\documentclass{caosp}

\usepackage{graphicx}

\articleNo{195}
\pubyear{2010}
\volume{40}
\volnumber{1}
\firstpage{1}
\received{January 25, 2010}
\accepted{June 28, 2010}

\begin{document}

%
\hauthor{T.\,Krej\v{c}ov\'{a}, J.\,Budaj and V.\,Krushevska}

\title{Photometric observations of transiting extrasolar planet WASP\,-\,10\,b}

\author{
        T.\,Krej\v{c}ov\'{a} \inst{1,2} 
      \and 
        J.\,Budaj \inst{2}   
	\and
	  V.\,Krushevska \inst{2, 3}
       }

\institute{
           Masaryk University, 
           Department of Theoretical Physics and Astrophysics, 
           602\,00~Brno, The Czech Republic, \email{terak@physics.muni.cz}
         \and 
           {\lomnica}, \email{budaj@ta3.sk} 
	\and
	   Main Astronomical Observatory of 
           National Academy of Sciences of Ukraine, 
           27 Akademika Zabolotnoho St. 
           03680 Kyiv, Ukraine,  \email{vkrush@mao.kiev.ua}
 }

\date{January 25, 2010}

\maketitle

\begin{abstract}
Wasp\,-\,10\,b is a very interesting transiting extrasolar planet.
Although its transit is very deep, about 40\,mmag, there are very different
estimates of its radius in the literature.
We present new photometric observations of four complete transits of 
this planet. The whole event was detected for each transit and 
the final light curve consists of more than 1500 individual CCD exposures.
We determine the following system parameters: 
planet to star radius ratio $R_{\mathrm{p}}/R_{*}=0.168 \pm 0.001$, 
star radius to semimajor axis ratio $R_{*}/a = 0.094 \pm 0.001$ and
inclination $i=87.3 \pm 0.1$\,deg.
Assuming that the semimajor axis is $0.036\,9 \pm ^{0.0012}_{0.0014}$\,AU (Christian et al. 2009), we obtain
the following radius of the planet 
$R_{\mathrm{p}}=1.22 \pm 0.05\,R_{\mathrm{J}}$, 
and radius of the star $R_{*} = 0.75 \pm 0.03$. The errors include the uncertainty
in the stellar mass and the semimajor axis of the planet.
Surprisingly, our estimate of the planet radius is significantly
higher (by about 12\,percent) than the most recent value 
of Johnson et al. (2009).
We also improve the orbital period 
$P_{\mathrm{orb}}=3.092\,731 \pm 1\times 10^{-6}$\,days 
and estimate the average transit duration 
$T_{\mathrm{D}}=0.097\,4\pm 8\times 10^{-4}$\,days. 

\keywords{extrasolar planets -- transit -- light curve}
\end{abstract}

\section{Introduction}
\begin{figure}
\centerline{
 \includegraphics[width=6.0cm,clip=]{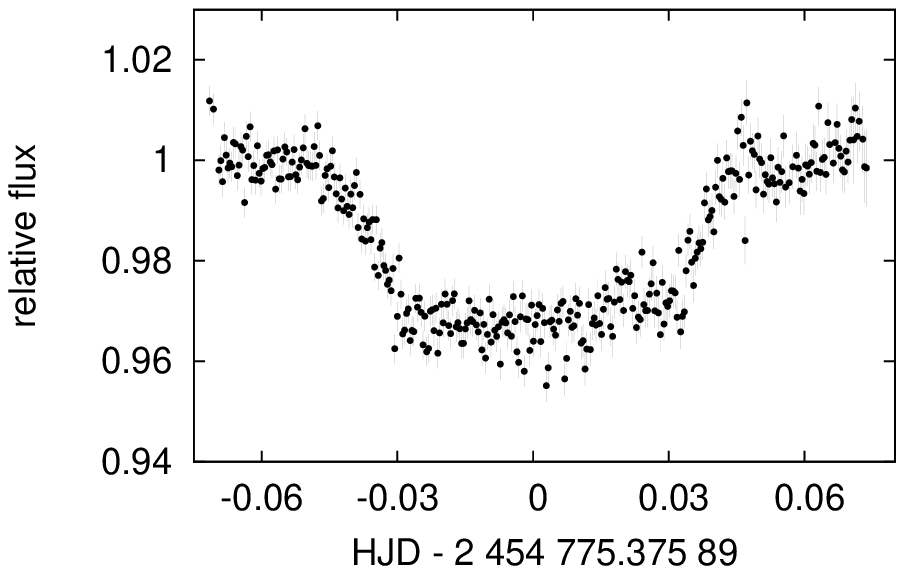}
 \includegraphics[width=6.0cm,clip=]{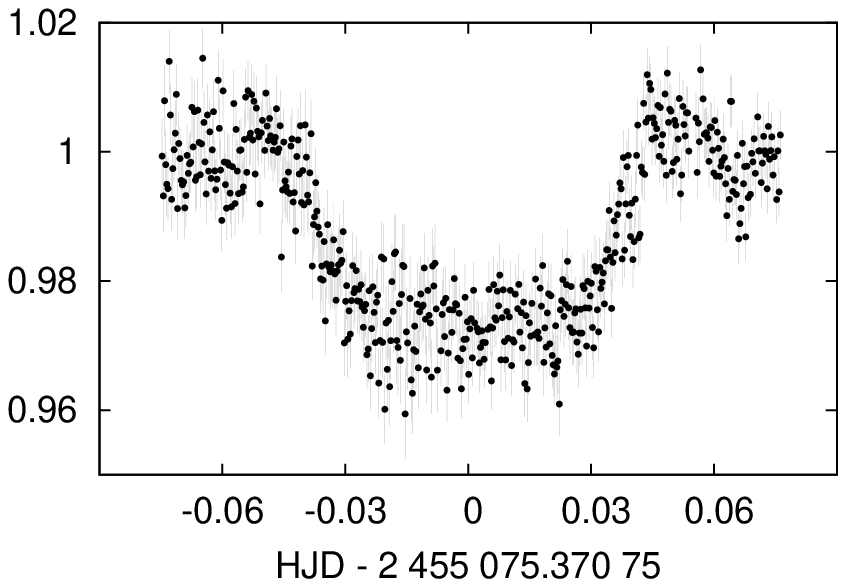}
}
\centerline{
 \includegraphics[width=6.0cm,clip=]{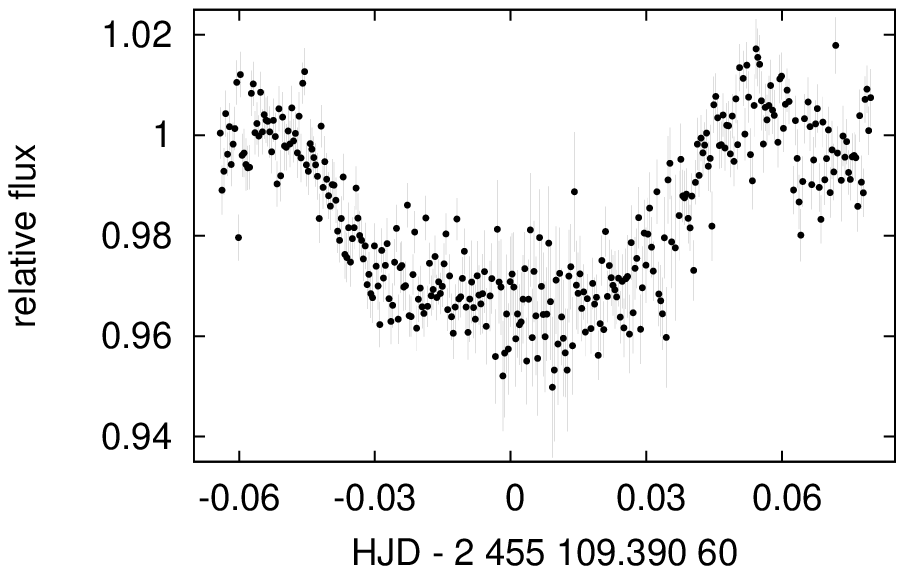}
 \includegraphics[width=6.0cm,clip=]{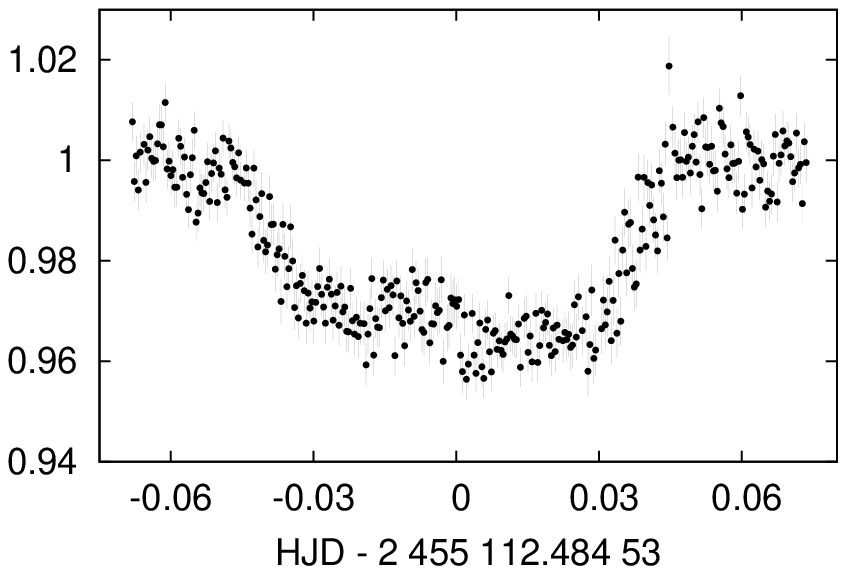}
}
\caption{The observations from the following nights from top left: 4/11/08, 31/8/09, 4/10/09 and 7/10/09 (dd/mm/yy).}
\label{f1}
\end{figure}

Transiting extrasolar planets are the VIP's (very important planets) 
of the extrasolar planet community (exoplanets), being the key to the understanding of the physics and various processes
in the interior and atmospheres of substellar objects. 
The orientation of their orbits in space with respect to the observer
is so fortunate that they allow us to observe the transit of the planet
in front of the parent star as a dip in the light curve. 
Photometric observations of these transits (in combination with
spectroscopy) provide us with important parameters of the system:
orbital period, inclination, mass and radius of the planet. 
The planet radius is a complicated function
of the planet mass, age, chemical composition as well as properties
of its orbit and the parent star
(Guillot \& Showman 2002, Burrows et al. 2007,
Fortney et al. 2007, Baraffe et al. 2008, Leconte et al. 2010).
Precise knowledge of planetary radii
is essential for further progress in the field.

Wasp\,-\,10\,b is one of the 80 so far known transiting extrasolar 
planets (Schneider, 1995). 
It was discovered in 2008 by the Wide Angle Search for Planets (WASP)
Consortium (Christian et al. 2009). The exoplanet orbits the parent 
star GSC 2752-00114 (spectral type K5) at the distance of
0.036\,9\,AU and has the mass of $2.96\,M_{\mathrm{J}}$ (Christian et al. 2009). Wasp\,-\,10\,b has one of the deepest transit depth,
cca 39\,mmag (Poddany et al. 2010). This results in the planet radius
of about $1.28\,R_{\mathrm{J}}$ (Christian et al. 2009).
This radius is quite large and an additional internal heat source
is required to understand such large radii.
Unfortunately, their most precise observations do not cover the whole
transit events.

Soon after the discovery the properties of the exoplanet were revisited
by Johnson et al. (2009). They obtained superb observations with 
the UH 2.2m telescope which covered the whole transit and significantly 
improved the precision of the radius determination. 
However, contrary to the discovery paper, 
they determined a very different radius of the planet: 
$R_{\mathrm{p}}=1.08\,R_{\mathrm{J}}$, which is 16 percent smaller 
than that given in the discovery paper. The origin of the difference was not 
entirely clear. Nevertheless, the new radius is consistent with 
previously published theoretical radii for irradiated Jovian planets.
Miller et al. (2009) could explain its new radius without any additional 
heat sources.

\section{Observations and data reduction}

Our observations were made with the Newton 508/2500\,mm telescope
with the CCD camera SBIG ST10 XME in R band (UBVRI system). 
The telescope is located in Star\'{a} Lesn\'{a} in the High Tatra Mountains
in Slovakia. 
The transiting system was observed during four nights
 --- 4/11/08, 31/8/09, 4/10/09 and 7/10/09 (dd/mm/yy). The time base for our 
observations was provided by the GPS device Motorola-Oncore M12+T. We performed the standard correction procedure (bias, dark and
flat field correction) and subsequently aperture photometry with
our data using the C-munipack software 
package\footnote{See \texttt{http://c-munipack.sourceforge.net}}. 
We carefully chose a few comparison stars and created one 
``artificial'' comparison star for each image as an average of individual
fluxes of all comparison stars. 
By means of such procedure we obtain a $1\sigma$ accuracy of about 3-5\,mmag
per 1 CCD exposure depending on observing conditions.

We also removed the linear trend in the out-of-transit data. 
The resulting light curves from the four nights are depicted in Figure \ref{f1}.

\section{Data analysis}

\begin{figure}
\centerline{\includegraphics[height=8.5cm]{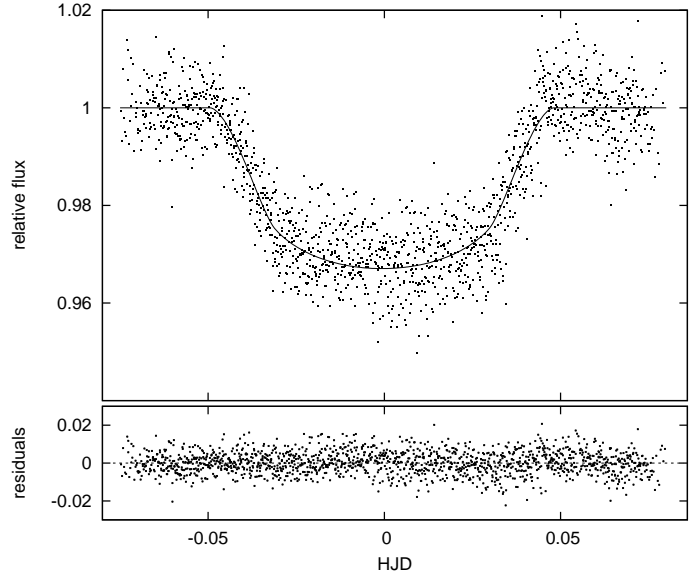}}
\caption{Top: Composition of 1\,533 data points from the four nights: 4/11/08, 31/8/09, 4/10/09 and 7/10/09 (dd/mm/yy) of the exoplanetary system Wasp\,-\,10\,b. The observation were made in filter R (UBVRI system). The solid line is the best fit transit curve. Bottom: Residuals from the best fit model.}
\label{f2}
\end{figure}
To obtain an analytical transit light curve we used the formulae from Mandel \& Agol (2002) assuming the quadratic limb darkening in the form:
\begin{equation}
I(\mu)/I(1) = 1 - c_{1}(1-\mu) - c_{2}(1-\mu)^{2},
\end{equation}
where $\mu$ is the cosine of the angle between the line of sight and the normal to the stellar surface, $I(\mu)$ is the intensity at the place defined by $\mu$, $I(1)$ is the intensity at the center of the disk and $c_{1}$ and $c_{2}$ are the coefficients of limb darkening. The limb darkening coefficients were linearly interpolated from Claret (2000) for the following star parameters: $T_{\mathrm{eff}}$\,=\,4\,675\,K, $\log(g)=4.5$ and $[M/H]=0.0$ (based on the results of Christian et al. (2009)). The coresponding coefficients for the R band are $c_{1}=0.5846$, $c_{2}=0.1689$.
For subsequent analysis we combined observations of all four transits into a single light curve. This final composition is displayed in Figure 2 and consists of more than 1\,500 individual CCD exposures.

To obtain the best fit parameters we found the minimal value of 
the $\chi ^{2}$ function given by:
\begin{equation}
\chi ^{2}=\sum _{i=1}^{N} \left(\frac{m_{i}-d_{i}}{\sigma_{i}}\right)^2,
\end{equation}
where $m_{i}$ is the model value and $d_{i}$ is the measured value of 
the flux, both for the $i^{\mathrm{th}}$ measured value; 
$\sigma_{i}$ is the uncertainty of the $i^{\mathrm{th}}$ measurement. 
For the minimization procedure we used the downhill simplex method 
(Press et al. 1992). We search for the optimal values of 
the following parameters: planet to star radius ratio 
$R_{\mathrm{p}}/R_{*}$, inclination $i$, 
the center of the transit $T_{\mathrm{c}}$ and 
the star radius to semimajor axis ratio $R_{*}/a$. 
The orbital period and limb darkening coefficients were fixed.

\begin{table}
\begin{center}
\footnotesize
\caption{Parameters of extrasolar system Wasp\,-\,10 from this work 
compared with the results from Christian et al. (2009) and 
Johnson et al. (2009).
$R_{\mathrm{p}}/R_{*}$ is the planet to star radius ratio,
$R_{*}/a$ is the star radius to semi major axis ratio,
$i$ is the inclination of the orbit,
$P_{\mathrm{orb}}$ is the orbital period,
$R_{\mathrm{p}}$ is the planet radius,
$R_{*}$ is the host star radius, and
$T_{\mathrm{D}}$ is the transit duration assuming 
the semimajor axis $a = 0.036\,9 \pm ^{0.0012}_{0.0014}$\,AU (Christian et al. 2009).}
\label{tabulka}
\begin{tabular}{lr@{\,$\pm$\,}lr@{}lr@{}l}
\hline \hline
Parameter & \multicolumn{2}{c}{This work} & \multicolumn{2}{c}{(Christian et al. 2009)} & \multicolumn{2}{c}{(Johnson et al. 2009)} \\
\hline
$R_{\mathrm{p}}/R_{*}$        & $0.168$ & $0.001$            & $0.170\,3$ & $\pm 0.002\,9$                 & $0.159\,18$ & $^{+0.000\,5}_{-0.001\,15}$  \\
$R_{*}/a$			&$0.094$ & $0.001$ 		& \multicolumn{2}{c}{---}			& $0.086$ & $\pm 0.009$	\\
$i$\,[deg]                    & $87.3$ & $0.1$               & $86.9$ & $^{+0.6}_{-0.5}$                 & $88.49$ & $^{+0.22}_{-0.17}$                   \\
$P_{\mathrm{orb}}$ [days]     & $3.092\,731$ & $1\times 10^{-6}$ & $3.092\,763\,6$ & $^{+0.000\,009\,4}_{-0.000\,021}$ & \multicolumn{2}{c}{---}\\
$R_{\mathrm{p}}$ $[R_{\mathrm{J}}]$ & $1.22$ & $0.05$        & $1.28$ & $^{+0.077}_{-0.091}$             & $1.08$ & $ \pm 0.02$  \\
$R_{*}$\,[$R_{\odot}$]        & $0.75$ &$0.03 $              & $0.775$ & $^{+0.043}_{-0.040}$            & $0.698$ & $ \pm 0.012$\\
$T_{\mathrm{D}}$ [days]       & $0.097\,4$ & $8\times 10^{-4}$  & $0.098\,181$ & $^{+0.001\,9}_{-0.001\,5}$       & $0.092\,796$ & $^{+0.000\,33}_{-0.000\,28}$\\
\hline \hline
\end{tabular}
\end{center}
\end{table}

To estimate the uncertainties of the calculated transit parameters 
we employed the Monte Carlo simulation method (Press et al. 1992). 
We produced about 1\,000 synthetic data sets with the same probability 
distribution as the residuals of the fit in Figure \ref{f2}. 
From each synthetic data set obtained this way we estimated 
the synthetic transit parameters. 
Subsequently, we were able to determine the uncertainties of 
the real parameters from the distribution of synthetic parameters. 
Finally, we took into account the uncertainty of the stellar mass and
semimajor axis according to the simple error accumulation
rule. The uncertainties of the planet and stellar radius are dominated 
by the uncertainties in the semimajor axis.
In case the value of the host star mass and/or the semimajor axis 
are revisited, our results and their errors can be simply rescaled. 

The orbital period $P_{\mathrm{orb}}$ was determined by 
the linear fit according to the following equation for the known time 
of central transits $T_{\mathrm{c}}(E)$ and the epoch $E$:
\begin{equation}
T_{\mathrm{c}}(E)=T_{\mathrm{c}}(0)+EP_{\mathrm{orb}},
\end{equation}
where $T_{\mathrm{c}}(E)$ is the central time for epoch $E$ and $T_{\mathrm{c}}(0)$ is the central time for $E=0$. We present (Table \ref{tabulka}) a preliminary result of the orbital period $P_{\mathrm{orb}}$. A more detailed analysis of the transit timing variations will be presented elsewhere. 




\section{Conclusions}

We have presented the observation and analysis of four light curves
of exoplanetary system Wasp\,-\,10\,b. 
This enabled us to improve the orbital period and revisit the properties 
of the system. The resulting values of the parameters
together with their uncertainties are given in Table \ref{tabulka}.
For comparison, the parameters from previous works are added there, too. 
Figure \ref{f2} shows the resulting best fit light curve together
with measured data from all four nights.

Our results indicate that the radius of the planet is significantly
larger than the latest estimate of Johnson et al. (2009).
Our observations are in agreement with the original observations
of Christian et al. (2009), but are more precise.
This is quite surprising since Johnson et al. (2009)
had seemingly superior observations\footnote{Recently, after the submission of our paper, new independent observations and analysis of Wasp\,-\,10\,b with similar conclusions were carried out by Dittmann et al. (2010).}.
Nevertheless, our value of the planet radius indicates that 
an additional alternative mechanism 
is required to inflate the planet size.
Enhanced opacities in the atmosphere (Burrows et al. 2007) or a tidal 
heating mechanism studied recently by Miller et al. (2009), 
Ibgui et al. (2010) and Leconte et al. (2010) might be invoked to explain 
the inflated radius of Wasp\,-\,10\,b.



\acknowledgements
We want to thank Dr. S.\,Shugarov for his help with the observations 
of exoplanetary transits, Dr. Gracjan Maciejewski, Dr. R.\,Kom\v{z}\'{i}k and Mgr. Marek Chrastina 
for fruitful discussion, and Dr. Pribulla and Dr. Hrudkov\'{a} for their
constructive comments on the manusctript. 
This work has been supported by 
the Marie Curie International Reintegration Grant FP7-200297, 
grant GA \v{C}R GD205/08/H005,
the National scholarship programme of Slovak Republic, and 
partly by VEGA 2/0078/10, VEGA 2/0074/09.

\end{document}